\documentclass[fleqn,usenatbib]{mnras}

\usepackage[T1]{fontenc}
\usepackage{tabularx}

\DeclareRobustCommand{\VAN}[3]{#2}
\let\VANthebibliography\thebibliography
\def\thebibliography{\DeclareRobustCommand{\VAN}[3]{##3}\VANthebibliography}

\usepackage{graphicx}	
\usepackage{amsmath}	
\usepackage{multicol}
\usepackage{amssymb}	

\usepackage[dvipsnames]{xcolor} 
\usepackage{newtxtext,newtxmath}

\title[The O3(a+b) GW transient mass distributions with \textsc{bpass}]{Forward Modelling the O3(a+b) GW transient mass distributions with \textsc{bpass} by varying compact remnant mass and SNe kick prescriptions}

\author[S. Ghodla et al.]{Sohan Ghodla$^{1}$, Wouter G. J. van Zeist$^{1}$, J. J. Eldridge$^{\thanks{j.eldridge@auckland.ac.nz}{1}}$, Héloïse F. Stevance$^{1}$,
    \newauthor
    Elizabeth R. Stanway$^{2}$
\\
$^{1}$Department of Physics, University of Auckland, Private Bag 92019, Auckland, New Zealand\\
$^{2}$Department of Physics, University of Warwick, Gibbet Hill Road, Coventry, CV4 7AL, UK
}

\date{Accepted XXX. Received YYY; in original form ZZZ}

\pubyear{2021}

\begin{document}
\label{firstpage}
\pagerange{\pageref{firstpage}--\pageref{lastpage}}
\maketitle


\begin{abstract}
We present forward modelling from the \textsc{bpass} code suite of the population of observed gravitational wave (GW) transients reported by the LIGO/VIRGO consortium (LVC) during their third observing run, O3(a+b). Specifically, we predict the expected chirp mass and mass ratio distributions for GW transients, taking account of detector sensitivity to determine how many events should have been detected by the current detector network in O3(a+b). We investigate how these predictions change by alternating between four different remnant mass estimation schemes and two supernovae (SNe) kick prescriptions. We find that none of the model populations resulting from these variations accurately match the whole O3(a+b) GW transient catalog. However, agreement from some models to part of the catalog suggests ways to achieve a more complete fit. These include reducing the number of low mass black holes (BHs) close to the mass gap, while also increasing the number of higher mass BHs below the pair-instability SN limit. Finally, we find that the interaction between the value of the remnant mass from a stellar model and the choice of SN kick is complex and different kick prescriptions may be required depending on whether a neutron star or BH is formed.
\end{abstract}




\section{Introduction}

In the span of only almost half a decade, the detection of GW transients has become almost routine. After LVC O3(a+b), there are now 90\footnote{Overall, 11 events from O1+O2, 44 events from O3a and 35 events from O3b} known GW transients events with a probability of being a real astrophysical signal ($p_{\rm astro}) > 50 \%$
\citep{2020arXiv201014527A, GWTC-3}. Each event in itself reveals much about the evolution of binary stars (e.g. for individual studies see \cite{2016PhRvL.116x1102A, 2017PhRvL.119p1101A, 2020ApJ...896L..44A} and for evolutionary implications for binaries, see \cite{2016MNRAS.462.3302E, 2016Natur.534..512B, 2017NatCo...814906S, 2017MNRAS.472.2422M, 2018MNRAS.481.4009V, 2019MNRAS.482..870E, 2021MNRAS.502.1925L, 2021arXiv210302608B} and references therein). However with a growing understanding of the distribution of masses, we are now beginning to constrain the population of GW transients both through parametric models
(e.g. \citealt{2019PhRvX...9c1040A, 2020arXiv201014527A, GWTC-3}) and using the technique of population synthesis to understand the physical process of stellar evolution that gives rise to the mass distribution (e.g. \cite{2008ApJS..174..223B, 2017NatCo...814906S, 2017PASA...34...58E, 2018MNRAS.474.2959G} and references therein).

While binary population synthesis is a powerful tool, there are many uncertainties in the physics incorporated in these models (e.g. \citealt{2017PASA...34....1D}) as well as in the prescriptions used to model the star formation history and chemical evolution of the Universe  (\citealt{2016Natur.534..512B, 2017MNRAS.472.2422M, 2019MNRAS.482.5012C, 2020MNRAS.493L...6T, 2021ApJ...907..110B,2021arXiv210302608B}). 
In particular, apart from the uncertainties in evolution of massive single stars (eg. mass-loss rate, rotation, overshooting, etc.) the understanding of the mass-transfer and common envelope phase in binaries is also incomplete (e.g., \citealt{2013A&ARv..21...59I} and references therein). In addition, though the delaytime distribution is believed to mostly lie within 0-4.5 Gyrs (e.g.  \citealt{2021arXiv210506491F}), a subset of GW mergers has also been found to come from very long time delays (e.g., \citealt{2016MNRAS.462.3302E, 2017MNRAS.472.2422M}). As such, the wide distribution of possible delaytime intervals between the formation of the progenitor stellar binary of a GW transient and the inspiral event introduces sensitivity to past star formation conditions.
Another required assumption is the remnant mass prescription used to determine the compact remnant mass once the stellar model reaches its final state, which often does not proceed all the way to core-collapse for numerical reasons. To this end, several remnant mass prescriptions with different features also exist in the literature,  
e.g. see, \cite{2020MNRAS.499.2803P}.  

In this letter, we utilize \textsc{bpass} (Binary Population and Spectral Synthesis) code results to forward model the population of GW transients resulting from their underlying stellar population up till their chirp event, specifically predicting the distribution of expected chirp masses and mass ratios for them during LVC O3(a+b). Section \ref{section: Methods etc} outlines the numerical model, highlighting the differences between this and earlier \textsc{bpass} work. Specifically, here we vary two key uncertainties in our analysis, (i) the nature of the kick that remnants receive while undergoing a SN explosion and (ii) the prescription for calculating the resulting remnant masses. Section \ref{section:Results} presents our results on the number of expected detections and the predicted GW transient object mass distribution. We then compare these to the LVC O3(a+b) catalog \citep{GWTC-3} to see which of the prescriptions produce a mass distribution that closely matches their results, presenting discussion and concluding remarks in section \ref{section:Discussion}.

\vspace{-0.6cm}
\section{Methods and Prescriptions, Simulations and Observations} \label{section: Methods etc}

Our aim here is to forward model the population of GW transients that may have been detected in O3(a+b) (henceforth O3), making predictions independent of the O3  observations, and explore some of the stellar evolutionary uncertainties that might impact these results. 
We use four different schemes to estimate the remnant masses resulting from stellar evolution and two different kicks to investigate how they collectively affect the populations of the predicted GW mergers.

\begin{figure}
    \centering
	\includegraphics[width=0.95\columnwidth]{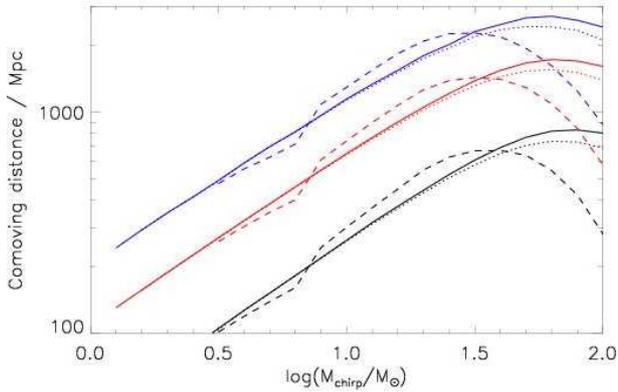}
	\vspace{-0.2cm}
  \caption{The detection distance versus chirp mass as a function of detection probability and mass ratio. 
    The solid, dotted and dashed lines represent the mass ratio values $q=1$, $q=0.5$ and $q=0.1$. The black lines are for a detection probability of 0.9, red 0.5 and blue 0.1.}
    \label{fig:detectionvolume2}
\end{figure}

\vspace{-0.45cm}
\subsection{Population synthesis}

We use \textsc{bpass} v2.2.1 stellar models \citep{2017PASA...34...58E,2018MNRAS.479...75S} and build upon our earlier work on predicting the GW transient rates. We refer the reader to those papers for full details of our model and method \citep{2016MNRAS.462.3302E,2019MNRAS.482..870E,2020MNRAS.493L...6T}. We use the same  underlying model of the cosmological star formation history and metallicity evolution as in \citet{2019MNRAS.482..870E}. However, two aspects of our previous population synthesis have been improved. 

Firstly, we have updated the treatment of rejuvenation during mass transfer onto the secondary stars. \textsc{bpass} has two principal sets of models: {\em primary models} where the more massive star's full structure is computed in detail and the secondary's evolution is approximated using the equations of \citet{2000MNRAS.315..543H}; and {\em secondary models} where the secondary is evolved in detail as either a single star or a star in orbit around a compact object -- depending on the outcome of the first SN. Many different primaries  map onto the same secondary model. As a result, in earlier \textsc{bpass} calculations we used an average rejuvenation age (i.e. time step at which mass transfer leads to rotational mixing, causing the accretor to become chemically homogeneous eg., \citealt{1987A&A...178..159M}) for each secondary model. In this work, we use a distribution of rejuvenation ages for each secondary model. This has the effect of smoothing out our delaytime distribution for transient events driven by GW emission.

Secondly, we now compute the GW merger time by calculating the orbital evolution in full as outlined by \cite{1964PhRv..136.1224P} rather than by simple interpolation between the two analytic forms for circular and highly eccentric orbits. This is motivated by the fact that many of the orbits lie in the mid-range between circular and eccentric after the second supernova, and the updated calculations offer a more accurate prediction of the merger times.

\vspace{-0.5cm}
\subsection{Detection horizons}
    

Detection of predicted mergers will depend on instrumental and alignment effects. Suites of codes to evaluate this do exist, such as \textsc{LALSuite} \citep{lalsuite}, but we implement our own Python package to do this because of our specific focus on population synthesis. The package \textsc{Riroriro}\footnote{A riroriro is a bird native to Aotearoa New Zealand that is frequently heard but seldom seen.} has been created with the aim of determining the fraction of predicted GW transients that can be detected \citep{riroriro}. Currently, it only deals with GW transients caused by the merger of NSNS, NS-BH, and BHBH. In the future, we plan to extend its scope to all areas of the GW population and the resulting GW spectral synthesis.

For each chirp mass and mass ratio pair \textsc{Riroriro} creates a synthetic merger gravitational waveform (including inspiral and ringdown phases). The effects of source orientation and distance are then taken into account to calculate an effective detection volume and the evolution of detection fraction of these events with luminosity distance. Our calculations span from 1~Mpc to 10~Gpc in steps of 0.1 dex. The package is inspired from \cite{buskirk2019} who created a \textsc{Mathematica} routine to calculate the expected GW waveform for a GW transient event. They used the results of \cite{huerta2017} to model the inspiral and ringdown emission. We use the same methods they outline  as well as using information from the \textsc{Findchirp} algorithm \citep{findchirp} to improve our modeling of the inspiral waveform and calculation of the signal-to-noise ratio (SNR). It is often difficult to match the inspiral and ringdown waveforms and we switch between them when the frequency and time derivative are closest between the two models.

For each resultant waveform, we use the method of \cite{barrett2018} to calculate the SNR of the merger model at a given redshift using the corresponding luminosity distance and redshifting the waveform. We subsequently calculate a total SNR assuming a triple-detection by combining the SNR calculated for the LIGO Hanford, LIGO Livingston, and VIRGO detectors during O3 in quadrature. We also calculate the average expected SNR of a GW observation with a single, non-specific detector operating. These SNR values assumed optimal alignment of detector and source. To take into account the effect of arbitrary orientation we multiplied these with a projection function \citep{projection1,projection2,projection3}, giving a probability distribution of the expected SNR and the probability that this would exceed the commonly-used detectability threshold value of 8, i.e. the detection probability. The variation in the detection probability of a GW transient depending on the chirp mass, mass ratio, and distance of the binary is illustrated in Fig. 
\ref{fig:detectionvolume2}. 
For each chirp mass, there is a maximum distance beyond which no mergers can be detected but there is a smoothly decreasing detection probability up to this horizon. Higher chirp masses can be detected to larger distances. However, at the highest chirp masses a hard upper limit is introduced due to the decreasing maximum frequency of these mergers. This upper limit is most sensitive to the mass ratio, as shown in Fig. \ref{fig:detectionvolume2}, and the lowest mass ratio has a significantly lower maximum chirp mass for the peak comoving distance to the source. We combine the detection probabilities as a function of chirp mass, mass ratio and distance with the population statistics from \textsc{bpass}, which dictate the frequency with which systems of such parameters form, to create a forward model for the LVC detection rates.
We integrate the rates in events per year per unit volume over the observable Universe to obtain rates in events per year. We use the method of \cite{hogg1999} to convert the luminosity distances used in the previous steps to comoving distances which are employed in the volume calculations\footnote{We assume a standard $\Lambda$CDM cosmology with $H_0= 100$h kms$^{-1}$ Mpc$^{-3}$ with h  = 0.696, $\Omega_\mathrm{M}$ = 0.286, $\Omega_\Lambda$ = 0.714}. We reduce our rates by a factor of (177.3 + 142.0) days / 365.25 days) to account for the elapsed duration during O3a + O3b where at least one detector was active and taking data. 

\vspace{-0.51cm}
\subsection{Remnant mass estimates and supernova kick velocities}


Within \textsc{bpass} the standard method of calculating remnant masses at the end of stellar evolution for massive stars is to compare the binding energy of the stellar envelope to the typical SN explosion energy of $10^{51}\,{\rm ergs}$. The ejecta mass is considered to be the amount of mass with this much binding energy, removed from the star, and the remainder goes into the remnant mass \citep{2004MNRAS.353...87E}. Here we refer to this method as the \textit{Standard} remnant mass scheme. However, to explore the effect of varying the remnant mass prescription on our predictions, we look into three other schemes.

(i) In \textit{AlwaysNS} scheme, motivated by the Chandrasekhar mass limit (and to save computation time), we force all the models to form a neutron star (NS) of 1.4~M$_{\odot}$ at core-collapse. Forcing all the remnants to form a NS at birth (albeit of the same mass) allows us to observe the influence this could have on the NS mergers rates. (ii) For most stars the final mass of the carbon-oxygen (CO) core would be a fair estimate of the maximum possible remnant mass, hence in M$_{\mathrm{CO, final}}$ remnant mass scheme, the final CO core mass of the progenitor model is set equal to the remnant mass (i.e. only the hydrogen/helium envelope will be ejected, independent of SN energy). (iii) For comparative purpose we employ a commonly-used rapid synthesis scheme from \citet{2012ApJ...749...91F}, here referred as \textit{FryerRapid}. This scheme gives a remnant mass that depends on M$_{\mathrm{CO, final}}$ (and partly on M$_{\mathrm{final}}$) of the stellar model. One interesting feature is that it predicts a mass-gap in the BH mass distribution with no BHs less massive than 5~M$_{\odot}$. Consequently, using the \textit{FryerRapid} scheme allows us to determine the impact of allowing for a mass-gap. Finally, in addition to using the above remnant mass prescriptions, we also use two different SN kick schemes namely the \cite{hobbs2005statistical} and \citet{bray2018neutron} kick velocity relations. The former is described by a Maxwell-Boltzmann distribution with 1-dimensional root mean square speed of 265 kms$^{-1}$ while the latter depends strongly by construction on the masses of the ejecta and the remnant and thus we expect the kick and remnant mass distributions to interact in a complex and non-linear way. Mathematically, the \textit{Bray} kick relation is as follows

\vspace{-0.3cm}
\begin{equation}
    v_{\mathrm{kick}}=100\left(\frac{M_{\mathrm{ejecta}}}{M_{\mathrm{remnant}}}\right)- 170\left( \frac{1.4}{M_{\mathrm{remnant}}}   \right) \label{2eq kick}
\end{equation} where $v_{\rm kick}$ is the velocity, in km s$^{- 1 }$, imparted to the compact remnant as a result of the SN, $M_{\mathrm{ejecta}}$ is the SN ejecta mass and $M_{\mathrm{remnant}}$ is the remnant mass (all masses in [M$_{\odot}$] units).

\vspace{-0.6cm}
\section{Results} \label{section:Results}

Below we show some ways in which one could constrain or verify the form of the SNe kick and the remnant mass relation using the GW transient data from O3.

\begin{figure*}
  \vspace{-0.5cm}
 \includegraphics[width=1.65\columnwidth]{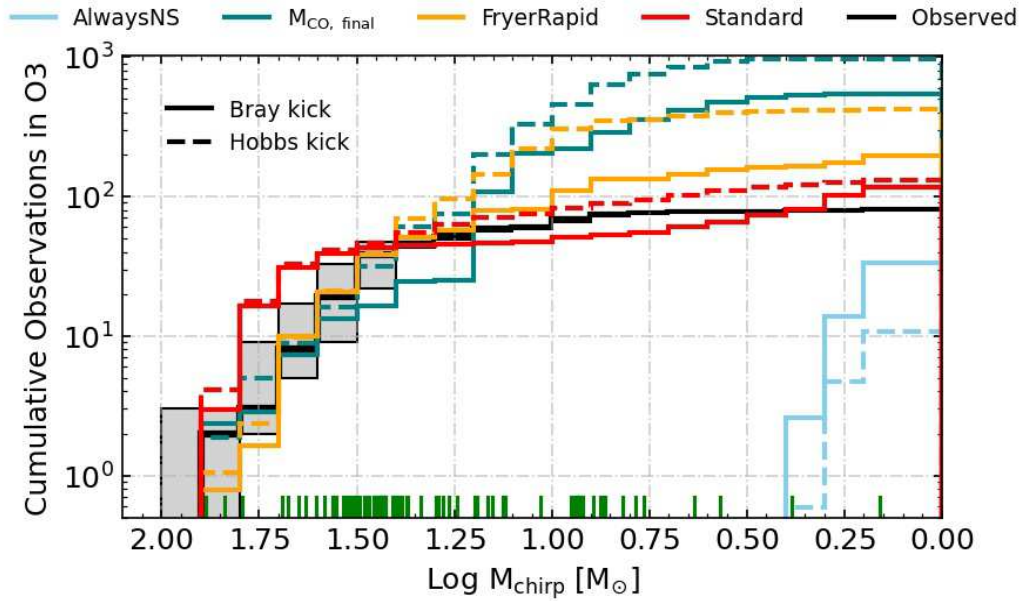}
 \caption[width=1\columnwidth]{The Chirp mass distributions of cumulative events observed by LVC O3(a+b) (black line) and those predicted by various schemes under the \textsc{bpass} model populations for the same duration of observation. The green ticks show the observed values of M$_{\mathrm{chirp}}$ and the solid grey area represents the uncertainty between the upper and lower M$_{\mathrm{chirp}}$ bounds for these observed events (\citealt{2020arXiv201014527A, GWTC-3}). The black curve is the cumulative observed number of events over the observed values of M$_{\mathrm{chirp}}$. Solid lines represent \textit{Bray} kick while dashed lines \textit{Hobbs} kick. }
  \label{fig:mchirp}
\end{figure*}

\subsection{Some implications based on merger rates}

\begin{table}
	\centering
    \caption{ Predicted number of observable detections during O3(a+b) for the three expected mergers types, employing  various methods for calculating  SN kicks and remnant masses. Rates in parenthesis are local intrinsic merger rates (in units of events Gpc$^{-3}$ yr$^{-1}$) calculated before the consideration of detection probabilities. Last row states the LVC O3(a+b)  observed and intrinsic rates. \citep{2020arXiv201014527A, GWTC-3}. The detectable rates account for the duration of 177.3 + 142.0 days of LVC O3 during which at least one detector was active.}
    \label{tab:mergerrates}

    \begin{tabular}{ccccc} 
    \hline
    
     Kick  &  Remnant  & NSNS 	& BHNS & BHBH \\
    \hline

    Hobbs & Standard &     4.9 (208) &    12.6 (131) &   45 (31) \\
    Hobbs & M$_{\mathrm{CO, \; final}}$ &  1.6 (43) &     46.8 (417) & 420 (873)\\
    Hobbs & FryerRapid &   3.8 (176)&    27 (208)&  171 (169)  \\
    Hobbs & AlwaysNS &    4.9 (223) &   0.4 (8.7)&   0.0  \\
    \hline
    
    Bray & Standard & 17.1 (745) &    13.3 (180) &  27.0 (13) \\
    Bray & M$_{\mathrm{CO, \, final}}$ & 5.6 (179)&   43.2 (498)&   212.5 (569) \\
    Bray & FryerRapid &      13.7 (677)&   16.0 (157) &    64.8 (70)\\
    Bray & AlwaysNS & 14.4 (708) &    2 (48)&   0.0 \\
    
	\hline\hline
	 \multicolumn{2}{l}{\textbf{LVC O3(a+b)} } &   1$\pm$1  &  5$\pm$2.2   & 69$\pm$8.3  \\
	 
	  \multicolumn{2}{l}{\textbf{Intrinsic [Gpc$^{-3}$  yr$^{-1}$]}}  & (13--1900)  &    (7.4--320)     &   (17.3--45)  \\
	 
    \hline
    \end{tabular}


\end{table}


Our results shown in Table \ref{tab:mergerrates} (also see Fig. \ref{fig:Visualise_Table_in_plot}, \ref{fig:Visualise_Table_in_plot2} in the Appendix), presents the expected number of events that should have been detected in O3 along with the predicted local intrinsic rates under each set of prescriptions. On considering these under the light of O3 merger rate data, we observe patterns with interesting implications:

\vspace{-0.2cm}
\begin{enumerate}
    \item The \textit{Bray kick} consistently gives a low BHBH intrinsic and detectable merger rate, as compared to the \textit{Hobbs kick}. The strength of these differences varies between the remnant mass schemes but a general trend holds. This is acceptable as in the \textit{Bray kick}, v$_{\rm kick}$ is inversely proportional to the M$_{\rm remnant}$. Thus more massive remnants receive weaker kicks and so have less eccentric orbits, longer merger times and a lower BHBH merger rate. In particular, we find that for very massive BHs the eccentricity of the orbit is independent of the kick but for intermediate mass BHs, \textit{Hobbs kick} produces more eccentric orbits and therefore more mergers. This behavior can be seen in Fig. \ref{fig:mchirp} where the cumulative observable count for both kicks closely follow each other up till $\sim$ 1.35 ($\sim$ 22 M$_{\odot}$) from the left, after which \textit{Hobbs kick} starts to produce more observable detection for the BBH systems.
    


    
    \item In M$_{\rm CO, final}$, the BHBH and BHNS intrinsic and detected rates are significantly higher than those seen in O3 implying that under our current settings, stellar remnants must be consistently less massive than the M$_{\rm CO, final}$ of their progenitors. Fig. \ref{fig:COcoremass2} (discussed later) also demonstrates this bias of M$_{\rm CO, final}$ toward higher mass remnants.

   
    \item The \textit{Standard} scheme over-predicts the number of detectable NSNS and BHNS mergers while under-predicting the BHBH mergers (fig \ref{fig:Visualise_Table_in_plot}), more so using the \textit{Bray kick}. On the other hand, a better agreement is seen once their intrinsic merger rates are compared with the expected local intrinsic rate. Additionally, the O3 BHBH merger count is reasonably reproduced by the \textit{Bray kick} + \textit{FryerRapid} remnant scheme. However, this produces too many mergers involving binary NSs. Though none of the schemes fully satisfy either the detectability or intrinsic criteria set by LVC O3, the \textit{Standard} + \textit{Hobbs} and \textit{Fryer Rapid + Bray} offer the closest fit to the data in Table \ref{tab:mergerrates}. This can also be seen in figure \ref{fig:Powerlaw} which provides the local BHBH merger rate density distribution as a function of the primary BH’s mass. Both the \textit{Standard} + \textit{Hobbs} and \textit{Fryer Rapid + Bray} produce mergers owing only to the isolated binary evolution scenario and their rates mostly fall below the 90\% credibility interval. Other relations also show similar trends but the above two schemes along with \textit{Fryer Rapid + Hobbs} come the closest to the PowerLaw+Peak distribution. We note that though Fig. \ref{fig:Powerlaw} helps us to perform a check on the validity of the chosen schemes it is not yet possible to reach a definitive conclusion.
    

        
    
    \item The \textit{AlwaysNS} scheme does have some BHNS mergers despite no BHs being formed in core-collapse. These BHs are the result of the subsequent accretion-induced collapse (due to accretion from the companion) of the already formed NS. This gives an indication of the relative importance of such a channel compared to normal BHNS mergers where one of the BHs is formed directly at core-collapse. 
    
    
    
    \item On comparing Fig. \ref{fig:Visualise_Table_in_plot} and \ref{fig:Visualise_Table_in_plot2} we find that most NSNS and BHNS rates fail to fall within the allowed detectability range. On the other hand, a better agreement is seen once their intrinsic merger rates are compared with the expected local intrinsic rates (Fig. \ref{fig:Visualise_Table_in_plot2}). Also, we find that there are some variations in the relative magnitudes of the intrinsic rates in comparison to the detectable rates  for the same underlying remnant mass and SNe kick scheme. At present, we think that more detections (especially of NSNS and BHNS mergers) need to be performed before any substantive conclusion can be drawn from these trends.
    
\end{enumerate}

\subsection{Implications based on the chirp mass distribution}

\begin{figure}
	\includegraphics[width= 1\columnwidth, height = 6.5cm]{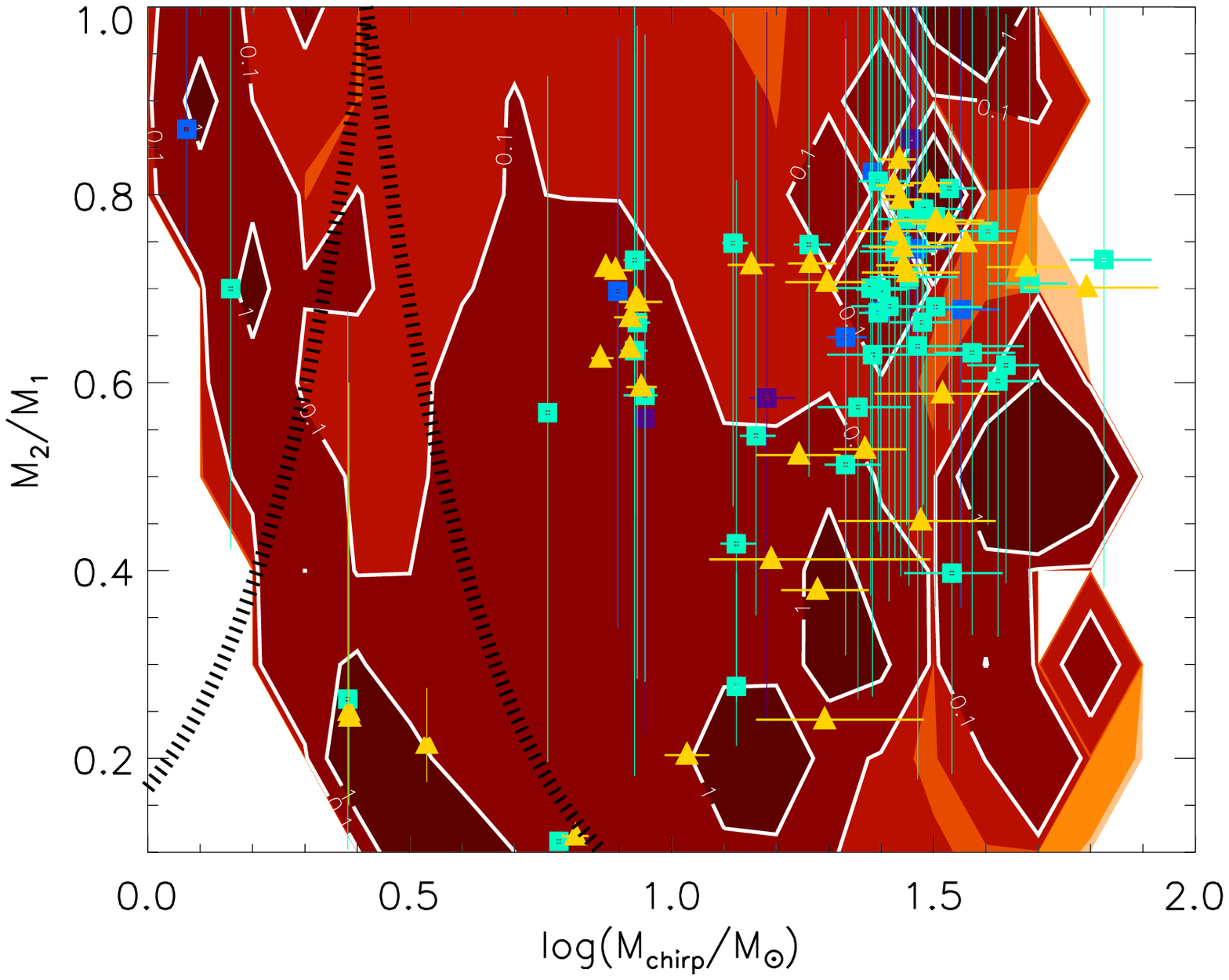}
	\includegraphics[width= 1\columnwidth, height = 6.5cm]{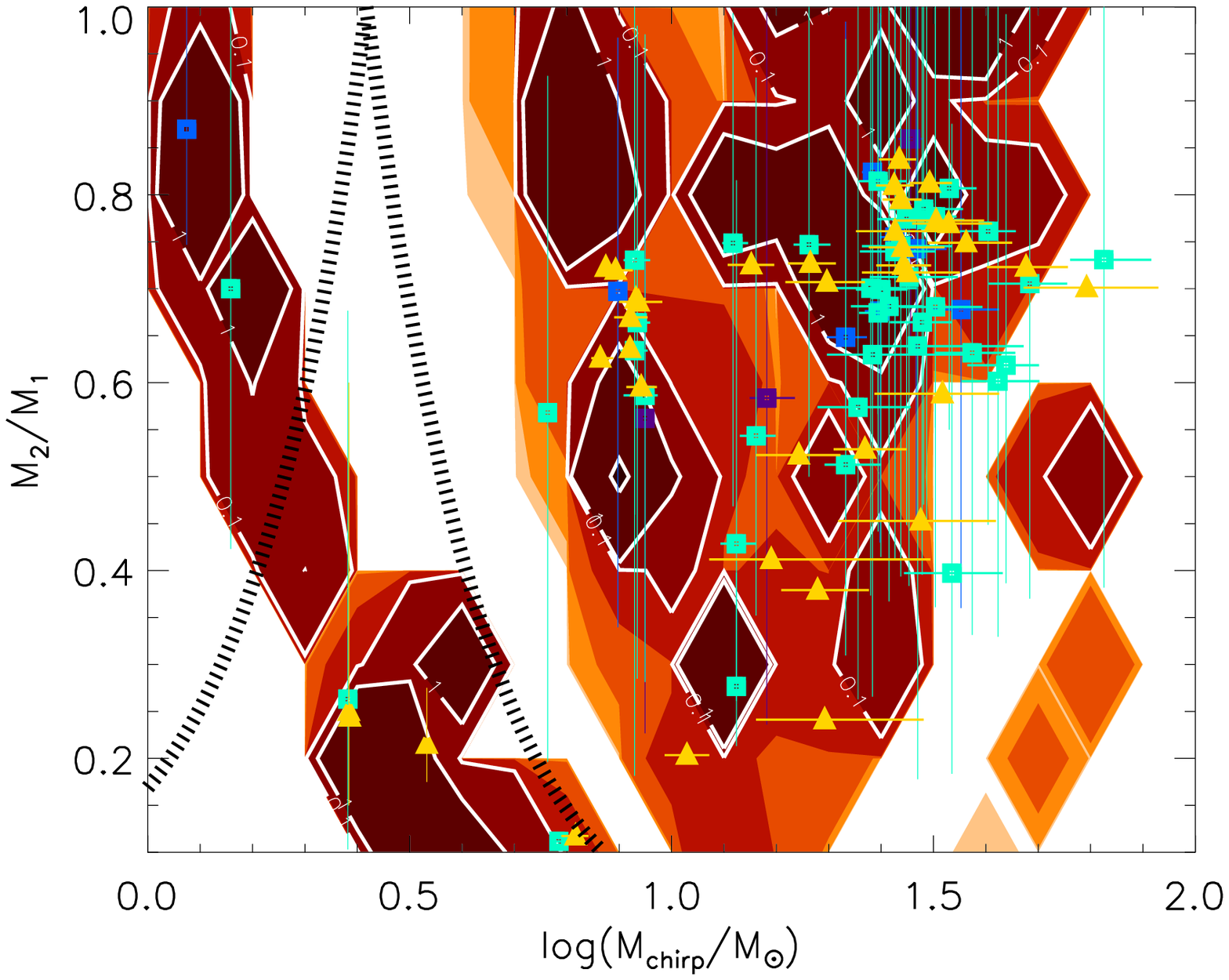} \vspace{-0.4cm}
   \caption{Contour plots of the expected distributions of detected events in O3 in M$_{\mathrm{chirp}}$ and mass ratio space. The upper panel shows the \textit{Hobbs kick} and the \textit{Standard} remnant mass scheme, the lower panel shows the \textit{Bray kick} and the \textit{FryerRapid} remnant mass scheme. Darker contours represent a higher probability. The black dotted line distinguishes the regions where mergers occur with component masses both either below 3M$_{\odot}$, one above and one below or both above  3M$_{\odot}$ (i.e, the expected regions of NSNS, BHNS, and BHBH mergers). The dark blue symbols are the observed events from O1, the blue symbols are from O2, the cyan symbols from O3a and the yellow triangles from O3b.}
    \label{fig:standard}
\end{figure}

Using rates alone is a limited approach to constrain the models. We must consider the number of events  expected in chirp mass and mass ratio space as well. To this end, we first consider the predicted chirp mass distribution of our model GW transient populations. Fig. \ref{fig:mchirp} shows the cumulative number of GW transients versus chirp mass from the high-mass end of the chirp mass distribution. We see that most of the model predictions reasonably match the observations at the high-mass end down to log 1.4 (i.e. $\sim$ 25~M$_{\odot}$), although the \textit{Standard} scheme (both kicks) tends to predict a higher count of massive mergers than observed. This can be seen more clearly in Fig. \ref{fig:Differential_rates} which provides the differential numbers for the same.  Below $\sim$ 25~M$_{\odot}$, the predicted distributions diverge significantly, with most schemes predicting more GW transients of lower chirp mass than observed. In the mass range of log 1.4 - log 0.5 ($\sim 25-5M_{\odot}$) \textit{Hobbs} kick produces more observable mergers but at the lowest chirp masses (spanning BHNS/NSNS mergers), populations arising from \textit{Bray kick} have an excess of systems. Cumulatively, the  \textit{Hobbs} kick tends to produce more detectable systems. However, among the schemes presented here, Fig. \ref{fig:mchirp} and \ref{fig:Differential_rates}  makes it possible to identify those that offer the closest match for the observed cumulative and differential events count over the M$_{\rm chirp}$ range.

\subsection{Implications based on chirp mass - mass ratio distribution}

We now further the analysis by comparing the GW transient distributions in chirp mass and mass ratio ($M_{\rm chirp}$ vs $q$) space. Fig.  \ref{fig:standard}, shows the schemes that offer the closest fit to the total event rates in Table \ref{tab:mergerrates} (i.e., \textit{Standard} + \textit{Hobbs kick} and \textit{FryerRapid} + \textit{Bray kick}). While the \textit{Standard} scheme spans nearly all combinations of remnant mass, however, in the \textit{FryerRapid} scheme the mass gap \citep{1998ApJ...499..367B, Ozel2010} becomes apparent due to a clear lack of BH systems near the BHNS/BHBH dividing lines. 

For log(M$_{\rm chirp}$/M$_{\odot}$) in the neighborhood of log 0.7 (i.e. $\sim$ 5~M$_{\odot}$) - close to the mass-gap end,  - the \textit{Standard} scheme overpredicts the number of expected mergers over most parts of the $q$ interval. The \textit{FryerRapid} scheme in Fig. \ref{fig:standard} does achieve a reasonably accurate peak near $M_{\rm chirp} \sim 1.5$ and $q \sim 0.7$.  However, it predicts a second, much larger peak for $M_{\rm chirp}$ and $q\sim 0.8$ and a stronger BHNS peak. In Fig. \ref{fig:standard} the peaks in the predicted and observed number of events for larger $M_{\rm chirp}$ lie at similar but not identical $M_{\rm chirp}$ vs $q$ ranges. This suggests a need for further investigation at the upper end of the BH mass distribution, close to the mass range for pair-instability SNe (e.g. \citealt{2017ApJ...836..244W, 2019ApJ...882..121S}). Additionally, in Fig. \ref{fig:standard2}, the \textit{Standard} + \textit{Bray kick} scheme underpredicts the BHBH and over-predicts the NSNS events, consistent with the findings in fig  \ref{fig:mchirp},  \ref{fig:Visualise_Table_in_plot} and  \ref{fig:Differential_rates}. The \textit{FryerRapid} + \textit{Hobbs kick} scheme in Fig. \ref{fig:standard2} also fails to provide a reliable match for the total events over the $M_{\rm chirp}$ vs $q$ range. The $M_{\rm CO, \; final}$ scheme in Fig. \ref{fig:COcoremass2} produces too many mergers nearly over the entire $q$ range at the lower and intermediate M$_{\rm chirp}$ values. In our simulation, these mergers could be avoided if the remnant mass is consistently less than $M_{\rm CO, \; final}$ in this $M_{\rm chirp}$ range which leads to a stronger kick and wider orbits. Overall, on comparing the distribution of the observed GW transients to our forward-model predictions, we find that the small number statistics still limit a meaningful comparison.


\section{Discussion \& Conclusions} \label{section:Discussion}

\vspace{-0.03cm}
Here, we have presented  theoretical predictions of the expected number of events (and their distribution in chirp mass and mass ratio space) detected within \textsc{bpass} under the LVC O3 sensitivity criteria. We find that none of the remnant mass schemes and SN kick combinations presented fully describe the observed population of GW transients in the O3 catalogs. Although most population models do reasonably well in reproducing the chirp mass distribution at masses above $\sim$ 25~M$_{\odot}$ but diverge considerably at lower masses. Our results suggest that the \textit{Standard} + \textit{Hobbs kick} scheme performs the best overall but requires modification at the upper and lower ends of the chirp mass distribution. The values of the most massive chirp masses predicted by this scheme need to be reduced by 0.1-0.2 dex, while the mass-gap needs to be included in some form at the lower end of the BH mass distribution. However, since the event GW190814 \citep{ 2020ApJ...896L..44A} is not predicted in the \textit{FryerRapid} scheme, this suggests that a prescription making low mass BHs significantly rarer, rather than preventing them from forming entirely within the mass-gap, may be required (see also, \citealt{Zevin2020}). 
The \textit{Standard} + \textit{Hobbs kick} somewhat underpredicts the observed BHBH events in Table \ref{tab:mergerrates}, (though it satisfies the intrinsic merger rate constraint) but these are the outcome of isolated binary evolution only (e.g. see, \citealt{Zevin2021}). We note that here we only vary the SNe remnant mass and SNe kick schemes while keeping the other uncertain parameters fixed (see, \citealt{2017PASA...34...58E}). Hence the predicted values have an implicit uncertainty built into them. Moreover, our models do not yet allow for tidally-induced chemical homogeneous evolution which (see \citealt{2009A&A...497..243D, 2016MNRAS.460.3545D, 2016A&A...588A..50M}) could result in a higher BHBH merger rate. Additionally, BHBH mergers could also come from dynamical channels in young stellar or globular clusters (e.g., \citealt{1993Natur.364..421K, 2000ApJ...528L..17P}) or at the disks of active galactic nuclei (e.g., \citealt{2012MNRAS.425..460M}) and some BHBH systems could be of primordial origin, arising in the early Universe \citep{1974MNRAS.168..399C}.
\citet{2020MNRAS.493L...6T} found that the GW transient rates involving BHs are highly dependent on the rate of chemical enrichment of the Universe and the number of BHNS and BHBH mergers can be reduced by altering the cosmic metallicity evolution. However, rapid enrichment would be required for our highest predicted rates to match those observed in O3. Lastly, all the predictions made here, assume fixed values for the rest of the uncertain parameters that go into the binary stellar evolution (see \citealt{2017PASA...34...58E} for more information). Changing these parameters could affect the results of this paper and hence any reliable prediction can only be made when more constraints (for the other unknowns) are available from other independent sources.

To conclude, we find some of our model populations produce a reasonable agreement between the predictions and observed distributions. Nonetheless, discrepancies remain, highlighting remaining uncertainties in our understanding of high mass compact binary formation. Our results indicate that no one model completely matches the observed distribution, although the \textit{Standard} remnant mass scheme + \textit{Hobbs kick} and \textit{FryerRapid} scheme + \textit{Bray kick} offer the closest match. The key features which are difficult to match for these model sets are the smaller number of low chirp mass transients and a significant peak around log 1.48 ($\sim$ 30~M$_{\odot}$). In the future, we might consider using a prescription where both the kick and remnant mass are stochastically sampled, similar to \cite{2020MNRAS.499.3214M}. Additionally, at the high mass end the physics of pair-instability and pulsational pair-instability systems also need to be taken into account (as in \citealt{2019ApJ...882..121S}). In Fig. \ref{fig:Visualise_Table_in_plot} and \ref{fig:Visualise_Table_in_plot2}, we find variations in the relative magnitudes of the intrinsic rates in comparison to the detectable rates for the same underlying remnant mass and SNe kick scheme. Given that supernovae remnant masses and kicks are interrelated, higher number statistics are needed to fully explore their effects. This should be possible as GW transient catalogs grow in size and completeness.


\vspace{-0.6cm}
\section*{Acknowledgements}

We thank the anonymous referee for their helpful feedback and suggestions. SG and WGJvZ acknowledge support from the University of Auckland. JJE and HFS acknowledge the support of the Marsden Fund Council managed through the Royal Society of New Zealand Te Apārangi. ERS received support from United Kingdom Science and Technology Facilities Council (STFC) grant
ST/T000406/1.


\vspace{-0.65cm}
\section*{Data Availability}


The data underlying this article will be shared on reasonable request to the corresponding author.




\vspace{-0.65cm}
\bibliographystyle{mnras}
\bibliography{refs} 




\appendix
\phantom{xxxx}

\vspace{-0.6cm}
\section{Supplementary Figures for GW event distribution}

\begin{figure*}
	\includegraphics[width=2\columnwidth]{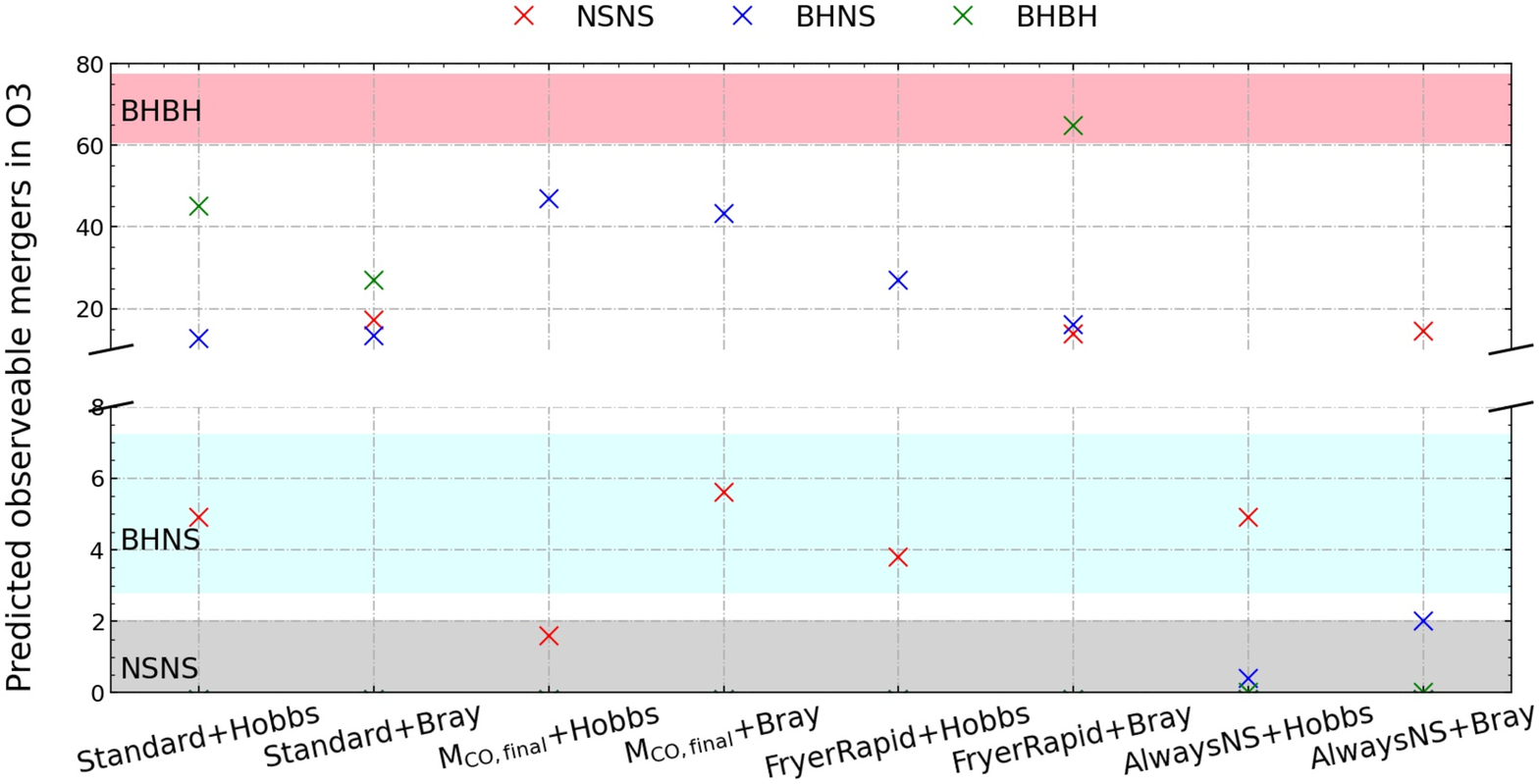}
   \caption{To assist in reading, here we plot the predicted number of detectable mergers in O3 as shown in Table \ref{tab:mergerrates} by the various schemes. Pink band represents the range of allowed values for BHBH, cyan for BHNS and grey for NSNS merger rates as obtained by LVC O3, \citep{2020arXiv201014527A, GWTC-3}. The red crosses are for NSNS, blue for BHNS and green for BHBH merger rates. The BHBH merger rate for M$_{\rm CO, final}$ (both kicks) and \textit{FryerRapid} (\textit{Hobbs kick}) are not shown due to their large values. Also, the BHBH merger rate for \textit{AlwaysNS} (both kicks) is zero.}
    \label{fig:Visualise_Table_in_plot}
\end{figure*}

\begin{figure*}
	\includegraphics[width=2\columnwidth]{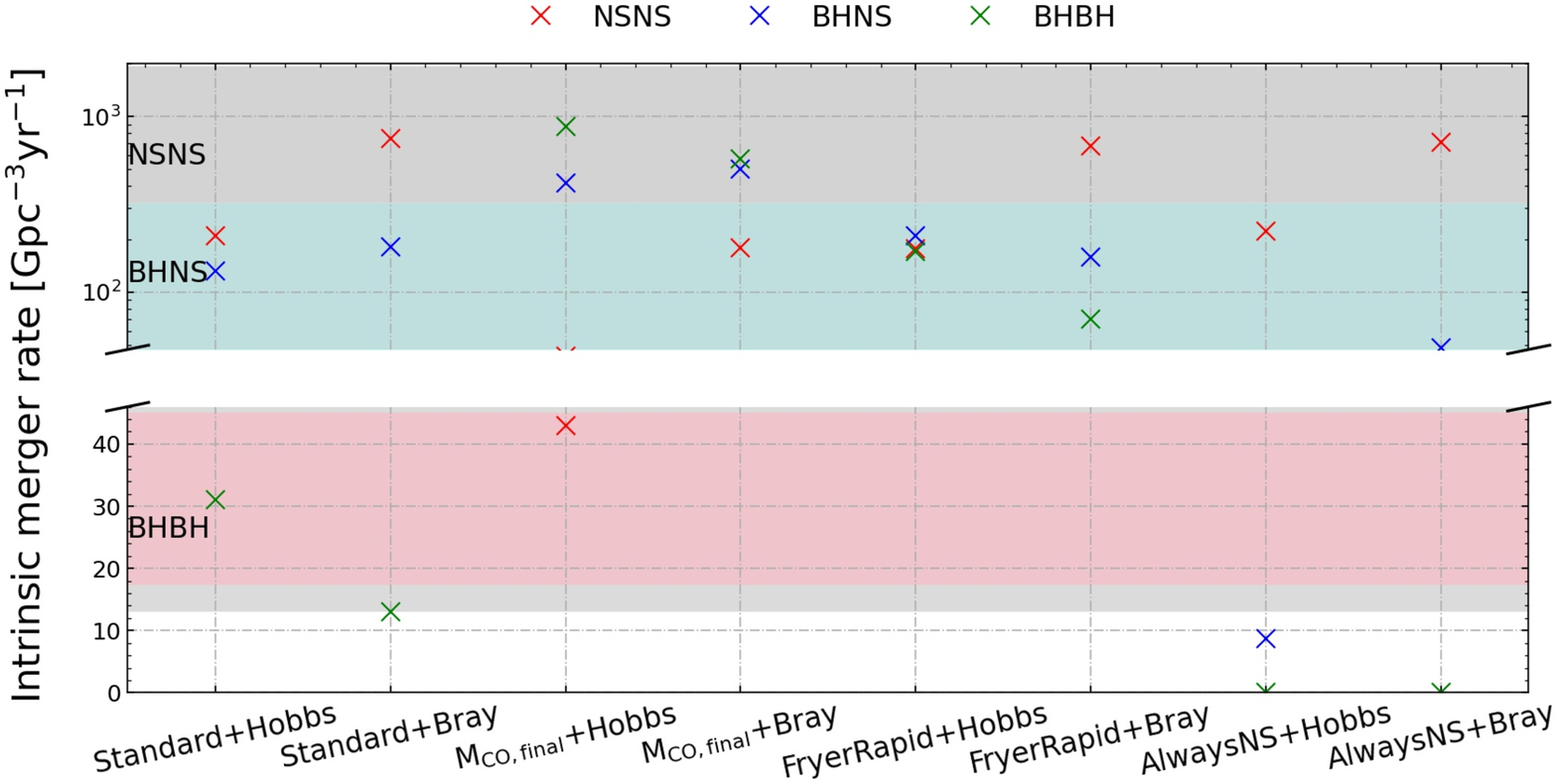}
   \caption{Here we plot the intrinsic local merger rates in Gpc$^{-3}$ yr$^{-1}$ as shown in Table \ref{tab:mergerrates} by the various schemes. Pink band represents the range of allowed values for BHBH, cyan for BHNS and grey for NSNS merger rates as obtained by LVC O3, \citep{2020arXiv201014527A, GWTC-3} using their assumptions for the underlying remnant mass distribution (which differs from those explored here). The overlapping cyan and grey region represents the overlap in the allowed ranges for the NSNS and BHNS mergers rates. The crosses have the same meaning. The BHBH merger rate for \textit{AlwaysNS} (both kicks) is zero. Note the change of scale on y-axis.}
    \label{fig:Visualise_Table_in_plot2}
\end{figure*}

\begin{figure*}
    \centering
    \includegraphics[width = 2\columnwidth]{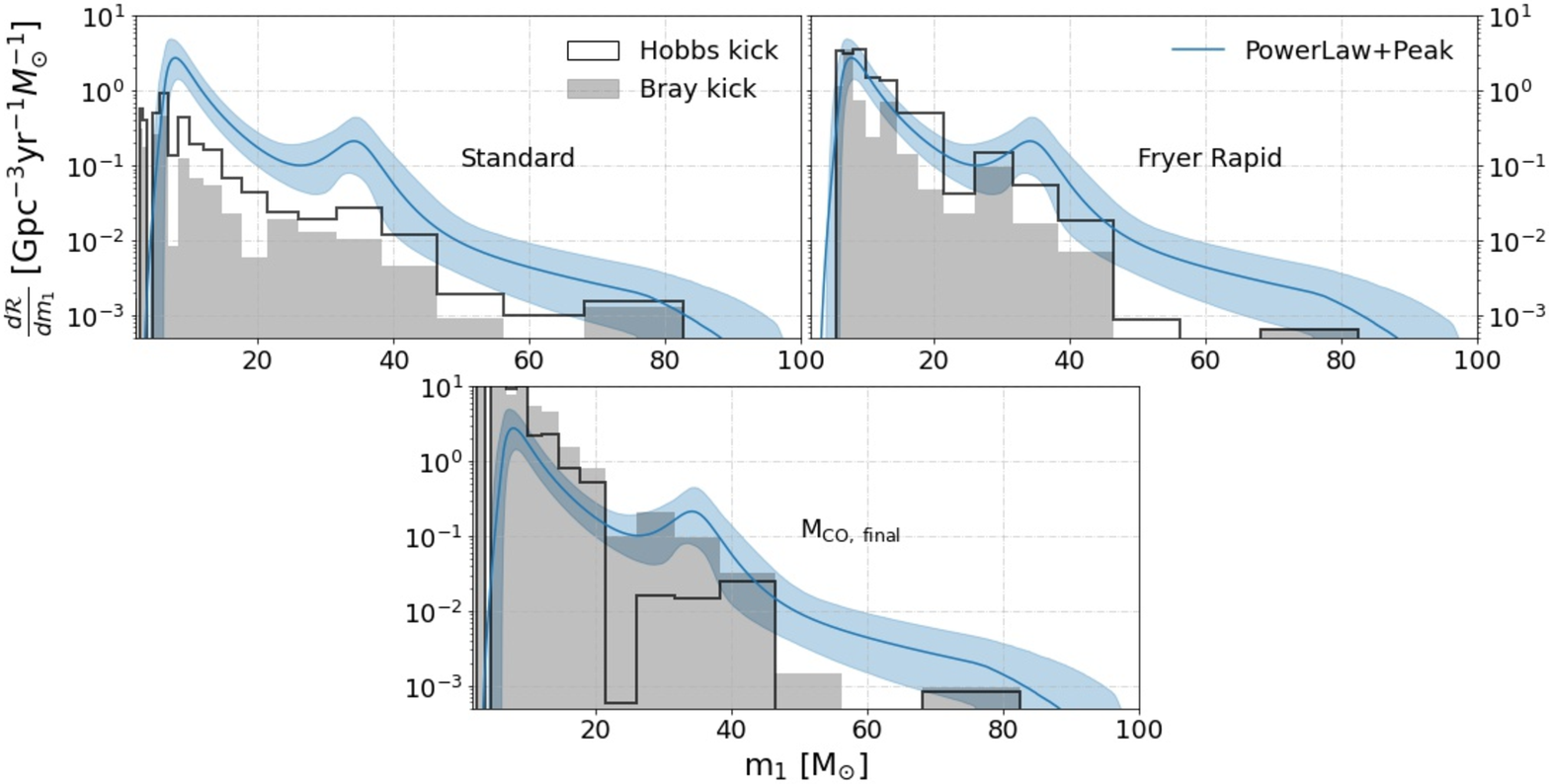}
    \caption{The local merger rate density (per unit solar mass) distribution  as a function of the primary BH's mass in the BBH mergers for the various schemes. The blue curve is the PowerLaw+Peak posterior population distribution for the same with its 90\% credible interval, represented by the shaded extension (the PowerLaw+Peak curve has been reproduced from \citealt{ZenodoLIGO} which is based on the work of \citealt{GWTC-3}).  The slope ($\alpha$) of the powerlaw in log-log space is  $\alpha=3.4_{-0.49}^{+0.58}$ supplemented with a Gaussian peak at $34_{-3.8}^{+2.3} M_{\odot}$. The unshaded histogram is for \textit{Hobbs} while the shaded is for \textit{Bray} kick respectively. The bins are logarithmically spaced and the minimum birth mass of BH is set to 2.5M$_{\odot}$ resulting in systems on the LHS of 3M$_{\odot}$ as well. The presence of systems in the pair-instability region is due to subsequent accretion by the BH from the companion star. Moreover, the data presented here does not yet account for the pulsational pair-instability SNe (e.g. \citealt{Woosley_PPISNe_2007}) and therefore does not feature a clear peak around 34 M$_{\odot}$.}
    \label{fig:Powerlaw}
\end{figure*}

\begin{figure*}
    \includegraphics[width=2\columnwidth]{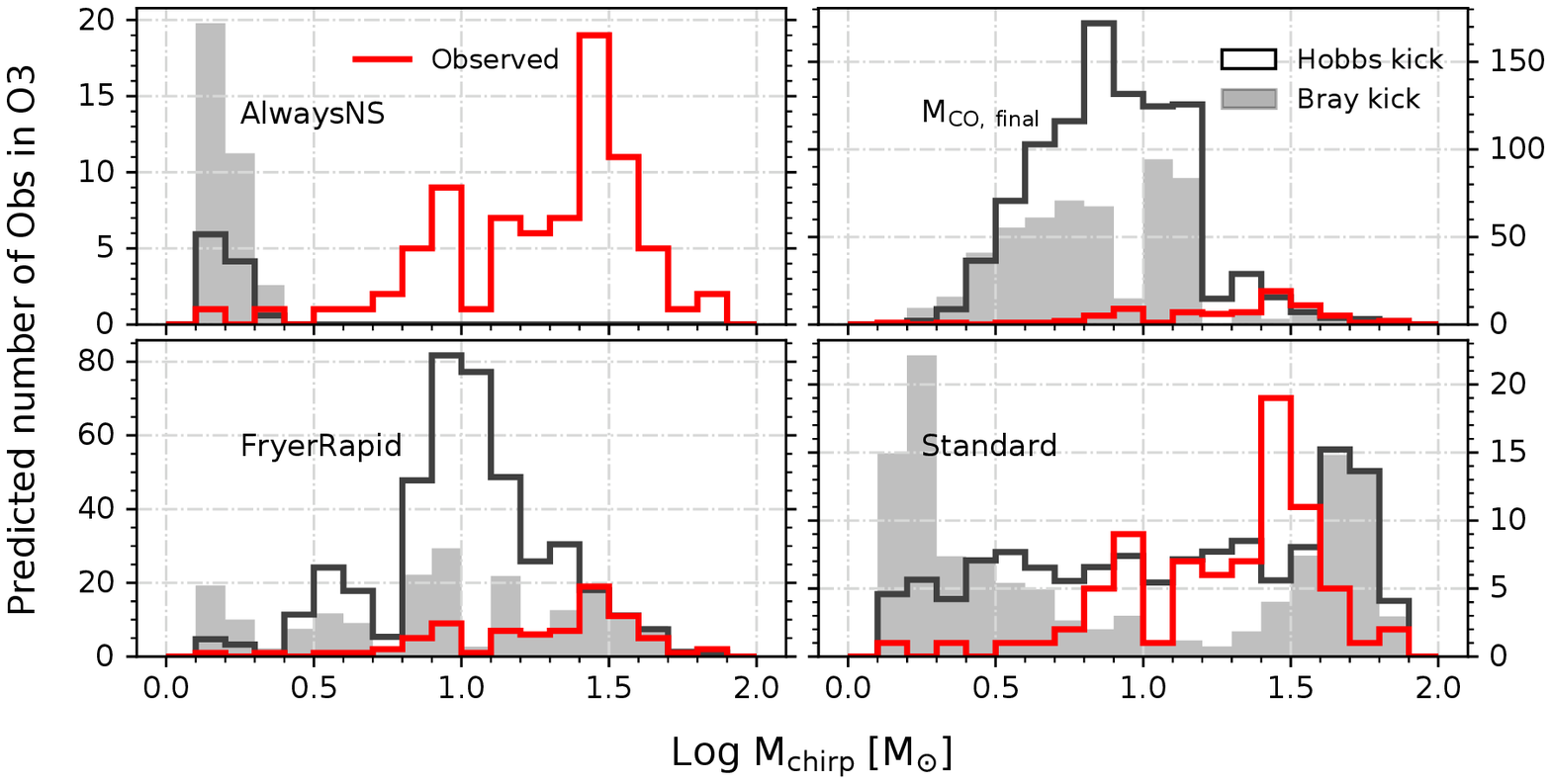}
  \caption{The predicted number of observations by various schemes as a function of chirp mass. The \textbf{red histogram} represents the number of observed events in O3 in their respective mass bin. The \textbf{filled histogram} represent \textit{Bray kick} while the empty ones represent \textit{Hobbs kick}. Note that the y-scale varies among the plots. Apart from the \textit{Standard + Hobbs} and \textit{FryerRapid + Bray}, others schemes predict large (or tiny) peaks at various M$_{\rm chirp}$ values that are highly inconsistent with the observed data. }
    \label{fig:Differential_rates}
\end{figure*}

\begin{figure*}
 	\includegraphics[width= 1\columnwidth, height = 7cm]{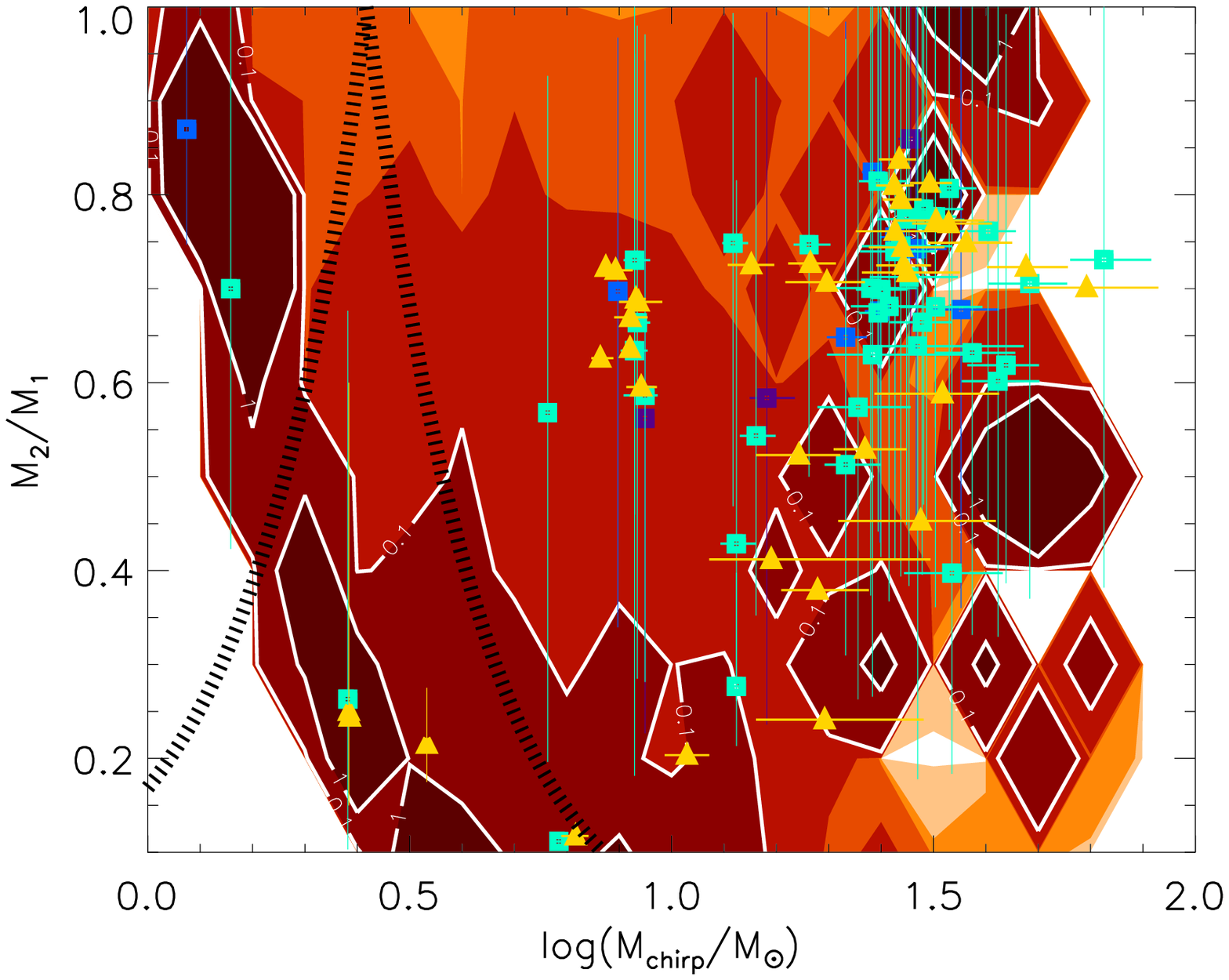}
 	\includegraphics[width= 1\columnwidth, height = 7cm]{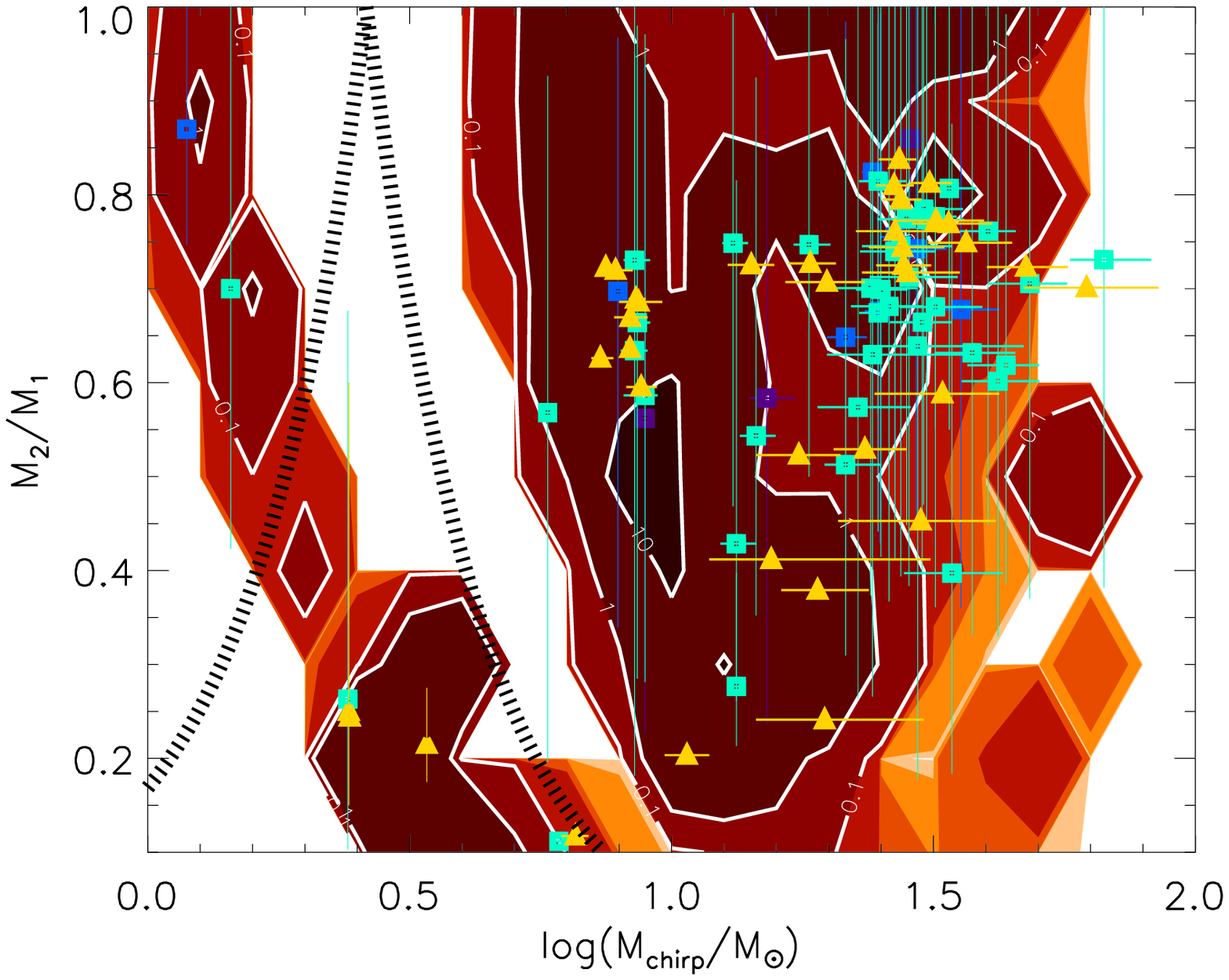}
   \caption{Details as in Fig. \ref{fig:standard} but here the LHS panel shows the \textit{Standard}  remnant mass prescription and the \textit{Bray} kick while the RHS shows the \textit{FryerRapid} remnant mass prescription and the \textit{Hobbs kick}}
    \label{fig:standard2}
\end{figure*}

\begin{figure*}
	\includegraphics[width=1\columnwidth, height = 7cm]{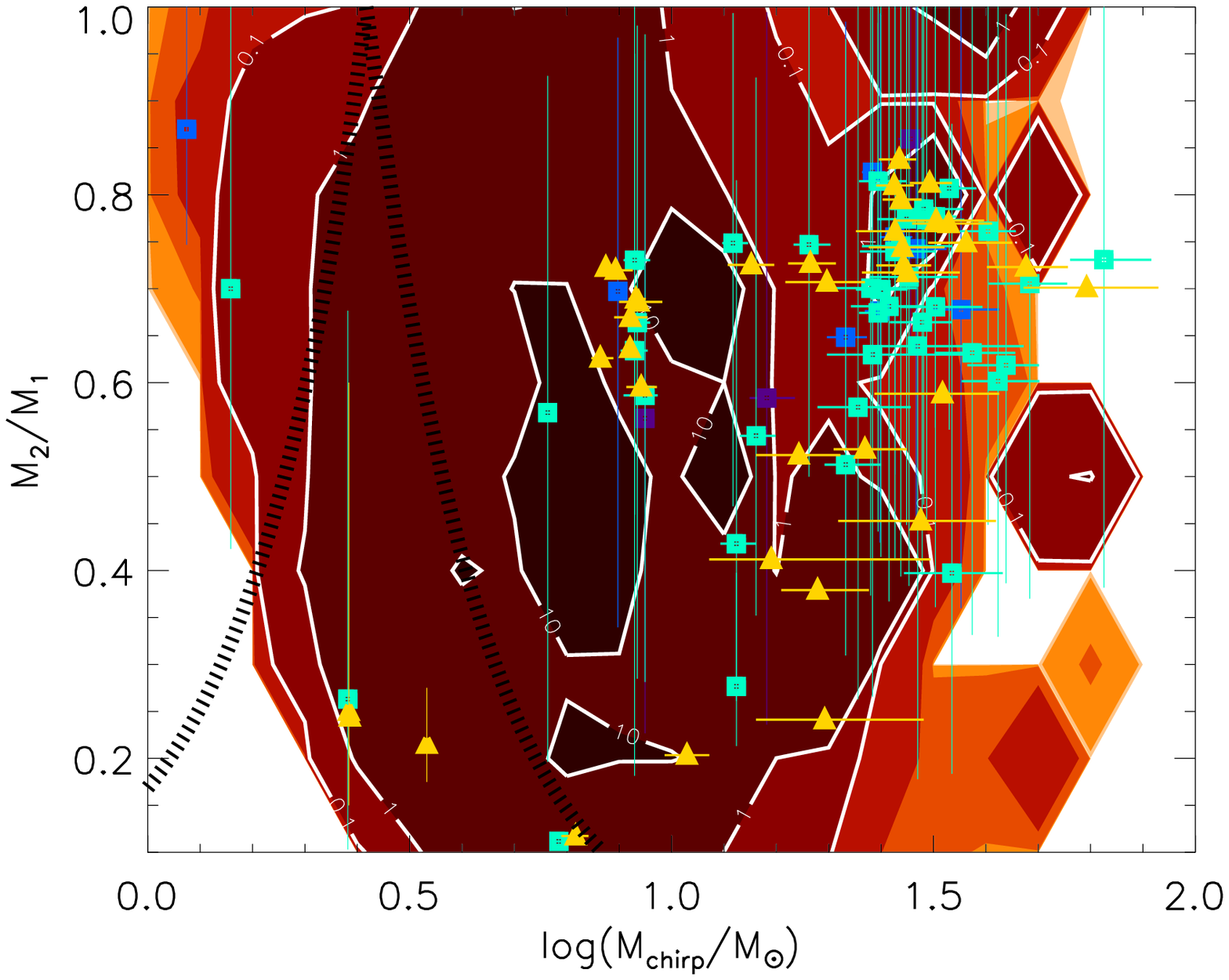}
 	\includegraphics[width=1\columnwidth, height = 7cm]{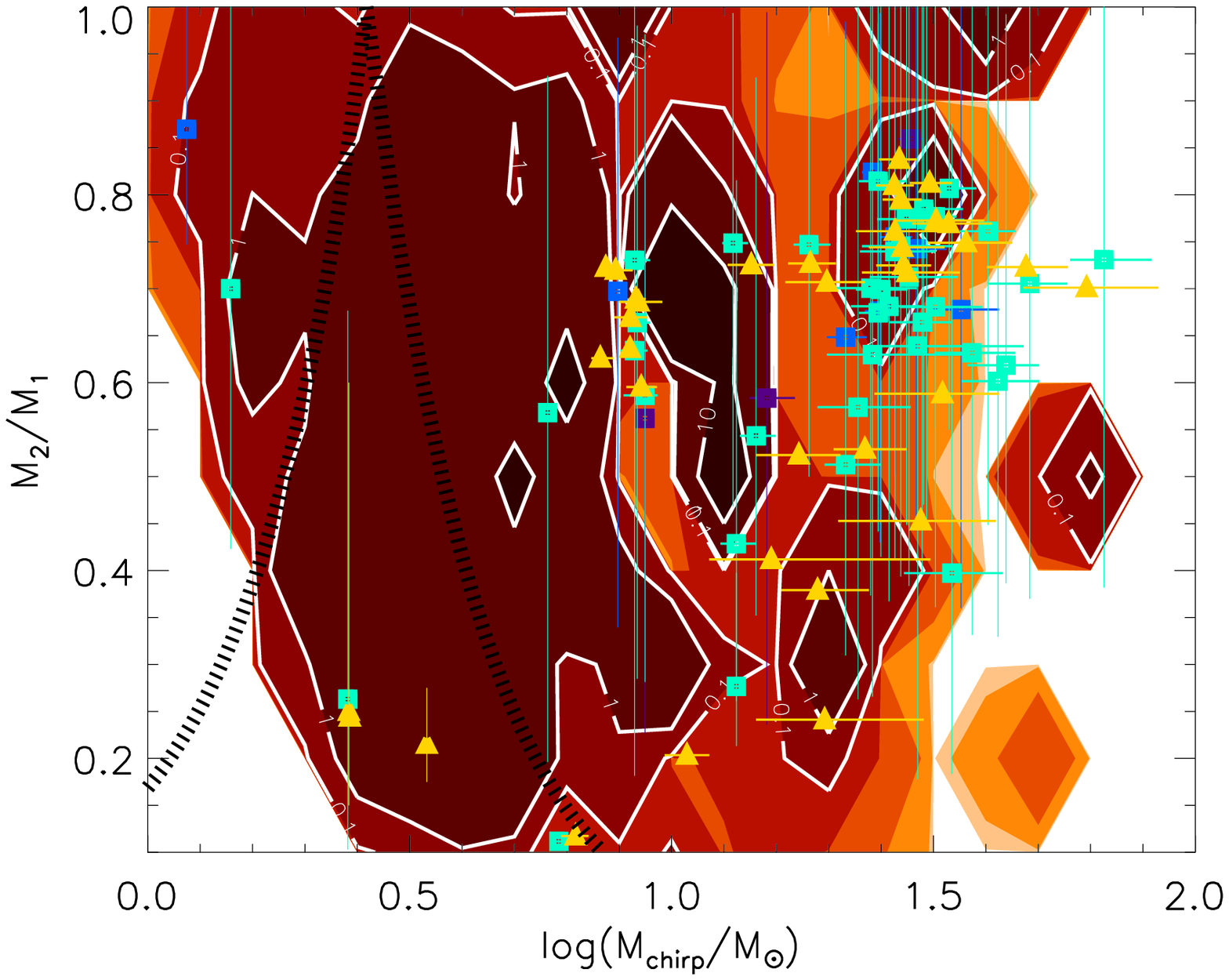}
   \caption{Details as in Fig. \ref{fig:standard}  but here the remnant mass is estimated from the M$_{\mathrm{CO, \; final}}$ remnant mass prescription with the LHS showing the \textit{Hobbs kick} and the RHS the \textit{Bray kick}. The mergers are biased towards the mid-range of the M$_{\rm chirp}$ values, providing a poor match to the observed data.}
    \label{fig:COcoremass2}
\end{figure*}

This appendix presents a visual plot (Fig. \ref{fig:Visualise_Table_in_plot}, \ref{fig:Visualise_Table_in_plot2}) of the data shown in Table \ref{tab:mergerrates}, the local merger rate density distribution as a function of the primary BH’s mass (Fig. \ref{fig:Powerlaw}) for the various schemes, the differential rates (Fig. \ref{fig:Differential_rates}) for data shown in Fig. \ref{fig:mchirp} and also the chirp mass - mass ratio distributions associated with the additional supernova kick and remnant mass prescriptions as discussed in the text (Fig. \ref{fig:standard2}, \ref{fig:COcoremass2}).


 
\bsp	
\label{lastpage}
\end{document}